\documentclass[10pt]{iopart}
\pdfoutput=1
\usepackage{graphicx}
\usepackage{cite}
\usepackage{color}

\begin{document}

\title[]{Impact of temporal scales and recurrent mobility patterns on the unfolding of epidemics}

\author{David Soriano-Pa\~nos$^{1,2}$, Gourab Ghoshal$^{3,4}$, Alex Arenas $^5$ and Jes\'us G\'omez-Garde\~nes$^{1,2}$}
\address{$^1$ Department of Condensed Matter Physics, University of Zaragoza, Zaragoza 50010, Spain.}
\address{$^2$ GOTHAM lab, Institute for Biocomputation and Physics of Complex Systems, University of Zaragoza, Zaragoza 50018, Spain.}
\address{$^3$ Department of Physics \& Astronomy, University of Rochester, Rochester, NY 14627, USA.}
\address{$^4$ Goergen Institute for Data Science, University of Rochester, Rochester, NY 14627, USA.}
\address{$^5$ Departament d'Enginyeria Inform\`atica i Matem\`atiques, Universitat Rovira i Virgili, Tarragona 43007, Spain.}
\vspace{10pt}
\begin{indented}
\item[] September 2019
\end{indented}

\begin{abstract}
Human mobility plays a key role on the transformation of local disease outbreaks into global pandemics. Thus, the inclusion of human movements into epidemic models has become mandatory for understanding current epidemic episodes and to design efficient prevention policies. Following this challenge, here we develop a Markovian framework which enables to address the impact of recurrent mobility patterns on the epidemic onset at different temporal scales. This formalism is validated by comparing their predictions with results from mechanistic simulations. The fair agreement between both theory and simulations enables to get an analytical expression for the epidemic threshold which captures the critical conditions triggering epidemic outbreaks. Finally, by performing an exhaustive analysis of this epidemic threshold, we reveal that the impact of tuning human mobility on the emergence of diseases is strongly affected by the temporal scales associated to both epidemiological and mobility processes.
\end{abstract}

%
%
%
%
%
\section{Introduction}

Despite the benefits associated with the development of human activity, globalization entails, as a counterpart, the acceleration of epidemic processes and its geographical expansion, turning local outbreaks into global epidemics \cite{humanactivity1}. Two of the main drivers of these changes in the global epidemic landscape are the explosion of human mobility through long-haul flights  \cite{mobility1,mobility2,mobility3,airport} or the effects of climate change \cite{climate1,climate2,climate3}. As a direct consequence of these two factors, during the last decades we have assisted  to the first reported autochthonous case of Dengue in Europe \cite{dengue1,dengue2} or the rapid expansion of Chikungunya virus across America \cite{chikun}. Thus, it is clear that the inclusion of human activity patterns, in particular those associated to mobility, in the formulation of epidemic models is mandatory for achieving accurate descriptions of actual epidemic scenarios.

Spurred by the relevance of human mobility patterns, many efforts have been devoted to determine the role played by human displacements in shaping the onset of epidemics. On the one hand, sophisticated agent-based models \cite{gleam1,gleam2,gleam3,gleam4} have been proposed to simulate real epidemic scenarios. These models have been very useful to unveil the most exposed areas to single epidemic outbreaks, thus enabling the design of policies to reduce their impact accordingly. Nevertheless, the extensive information required to address each particular case and the number of parameters involved in these modelsmake it difficult to obtain a clear interpretation about the impact of human mobility and its generalization to address different epidemic scenarios. On the other hand, for the sake of obtaining more general insights, theoretical frameworks have been also proposed to address the effects of human mobility on epidemic spreading. In this sense, the most extended way of incorporating human mobility patterns into epidemic modeling \cite{epi3,epi1,epi2} is the use of metapopulations \cite{metapop1,metapop2,metapop3,metapop4}. Metapopulations, originally introduced in the field of ecology, are networks whose nodes are identified with patches or areas across which a set of agents moves according to the flows dictated by the network topology. 

Theoretical works usually contain assumptions limiting their capability of fairly reproducing the spatio-temporal unfolding of epidemics. For this reason, epidemic models based on metapopulations have been continuously refined to bridge this gap between theory and agent-based models. Thus, first models considering agents as random walkers \cite{colizza1,colizza2,colizza3,colizza4} have been improved to include more realistic situations such as the recurrent nature of human mobility \cite{recurrent1,recurrent2,recurrent3,recurrent4,recurrent5,recurrent6} or the coexistence of different mobility patterns \cite{multmetapop1,multmetapop2}. In particular, recently we proposed a framework called MIR (Movement-Interaction-Return) model  \cite{natphys} which enables to introduce real data about the population distribution and their recurrent mobility patterns and to assess the impact of daily movements on the spread of diseases across a given city. One limitation of this model is that the time scales associated to both mobility and epidemiological processes are assumed to be equal, which implies that the duration of human movements is restricted to a single epidemiological step. 

The aim of this paper is to overcome this limitation by incorporating different time scales between epidemiological processes and mobility. For this purpose, we extend the original MIR model by proposing a generalized formalism which introduces a new parameter governing the time spent by agents outside their residence. The manuscript is organized as follows: In Section II, we describe the main features of the model and present the theoretical framework based on Markovian equations that we compare with those derived for the original MIR model. In Section III, we validate the former equations by comparing their theoretical predictions with results from mechanistic simulations. Once validated, we take advantage of them and we introduce in Section IV a theoretical estimation of the epidemic threshold. In Section V, we study how the recovery time and the typical staying time associated to movements change the dependence of the epidemic threshold on human mobility. Finally, in Section VI, we discuss the relevance of our findings and their implications for the design of prevention policies.

\section{The model}
\begin{figure}[t!]
\centering
\includegraphics[width=0.90\columnwidth]{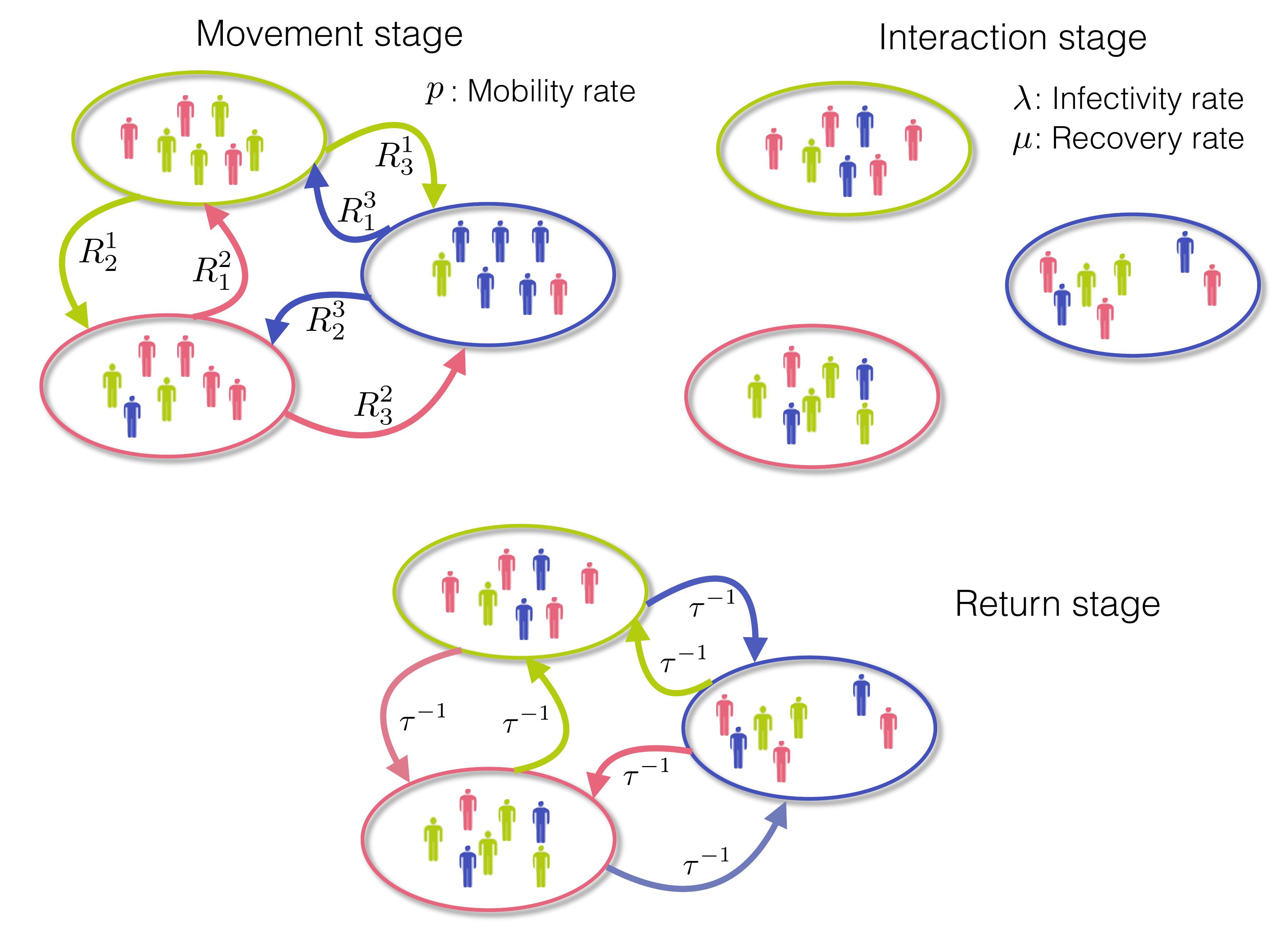}
\caption{Schematic representation of the generalized MIR model. Each time step involves three stages ${\it i)}$ Movement dictated by the degree of mobility $p$ and the mobility tensor ${\bf M}$, ${\it ii)}$ Interaction according to the contagious rate $\lambda$ and the recovery one $\mu$ and ${\it iii)}$ Return following a probability $\tau^{- 1}$. The arrows color code reflects the residence of the agents displaying each flow.}\label{fig1}
\end{figure}

\subsection{Model assumptions}
Let us start by presenting the extended version of the MIR model. In Fig.~\ref{fig1} we show a schematic representation of the microscopical processes contained in our model. Namely, we consider that every time step consists of three stages. In the first stage, agents are allowed to move across the network with a probability $p$. At this point, we must remark that, unlike the original MIR model, each patch can now be populated by individuals from different geographical areas. In this sense, given that residents and visitors usually display very different mobility patterns, we now must include an object governing human flows more complex than Origin-Destination (OD) matrices.  To do so, we introduce introduce a 3-dimensional tensor {\bf M} whose elements $M^{i}_{jk}$ tell us the probability that an agent belonging to node $i$ moves from patch $j$ to patch $k$ \cite{nota}. However, as we are only interested in strictly recurrent mobility patterns with some staying time at the destination, this tensor ${\bf M}$ can be decomposed into:

{\setlength{\mathindent}{0.5cm}

\begin{equation}
M^i_{jk}=\delta^i_{j}R^{j}_{k}+(1-\delta^{i}_{j})\delta^{j}_{k}\ .
\label{Eq.tensor}
\end{equation}
This choice imposes that only the individuals located at their residence are allowed to move to other neighboring areas. In particular, the first term indicates that these movements are governed by the commuting flows encoded in the OD matrix ${\bf R}$, being the element $R^{j}_{k}$ the probability that an agent associated to patch $j$ moves to patch $k$. In its turn, to include the staying times associated to human movements, visitors out of their residence are forced to stay at their temporal node until coming back home. Once all the movements have taken place, agents start to interact. For the sake of simplicity, we deal with Susceptible-Infected-Susceptible (SIS) diseases and assume the population to be well-mixed so all the agents interact with those placed in the same patch. This way, we define $\lambda$ as the probability that an infected agent transmits the pathogen to a susceptible one and $\mu$ as the probability that infected agents overcome the disease and become susceptible again. Finally, at the return stage, visitors decide whether staying out of their residence or to come back home with a probability $\tau^{-1}$, {\em i.e.}, $\tau$ accounts for the characteristic permanence time in the destination nodes. Note that $\tau^{-1}=1$ ($\tau=1$) recovers the original MIR model whereas, for $\tau^{-1}=0$, agents lose the information of their residence and are steadily redistributed across the neighboring areas.

\subsection{Model equations}
To build our formalism, we assume that the underlying metapopulation is divided into $P$ patches, each one with $N^i$ residents. The introduction of different temporal scales affecting mobility and the epidemiological processes requires to account for the temporal evolution of populations crowding each patch. In this sense, we denote as $n^i_j (t)$ the number of residents in $i$ who are visiting $j$ when time step $t$ starts. Following the aforementioned microscopical rules, the evolution of these quantities reads:
\begin{eqnarray}
n_j^i (t+1) &=& (1-\delta^{i}_{j})\left[(1-\tau^{-1})(pR^{i}_{j}n^i_i(t)+n^i_j(t))\right]\nonumber \\
&+&\delta^{i}_{j}\left[\left(1-p\left(1-R^{i}_{i}\right)\right)n^i_i(t)+\tau^{-1}\sum_{j\neq i}\left(pR^{i}_{j}n_i^i(t)+n^i_j(t)\right)\right].
\label{Eq.dynamicalpop}
\end{eqnarray}
The first line of the former expression contains the evolution of the distribution of agents outside their residence. These individuals are the sum of the agents who were already outside at time $t$ and decide not to come back and those moving from $i$ to $j$ at time $t$ and not returning to their residence. In its turn, the second line governs the temporal evolution of the residents in a geographical area. There, the first term encodes those agents who decide not to leave their residence during the movement stage whereas the second one contains  agents who are in a different place than their residence but come back there at the end of the time step. As long as the relaxation time for the spread of diseases is much longer than the mobility time scale ($\mu\ll\tau^{-1}$), one can assume that the distribution of agents across the metapopulation is given by the stationary solution of Eq. (\ref{Eq.dynamicalpop}), which can be expressed as:
\begin{equation}
n^i_j=(1-\delta^{i}_{j})\frac{N^iR^{i}_{j}}{1-R^{i}_{i}+\frac{\tau^{-1}}{p(1-\tau^{-1})}}+\delta^{i}_{j}\frac{N^i}{1+\frac{p(1-R^{i}_{i})(1-\tau^{-1})}{\tau^{-1}}}\,
\label{Eq.population}
\end{equation}

To track the evolution of SIS diseases, we need to account for the number of infected residents in $i$ visiting $j$ at each time step, denoted in the following as $\rho^i_j(t)$. This way, our formalism is composed by a set of $P\times E$ Markovian equations, where $E$ is the total number of connections of the underlying metapopulation. According to the assumptions of the model, the temporal evolution of $\rho^i_j(t)$ can be expressed as:
\begin{eqnarray}
\rho^i_j(t+1)&=&\delta^{i}_{j}\left[1-\mu\right]\left[\left(1-p\left(1-R^{i}_{i}\right)\right)\rho^i_i(t)+\tau^{-1}\sum_{l\neq j}\left(pR^{i}_{l}\rho_i^i(t)+\rho^i_l(t)\right)\right] \nonumber \\
&+&\delta^{i}_{j}\left(1-p\left(1-R^{i}_{i}\right)\right)\left(n^i_i-\rho^i_i(t)\right)\Pi_i(t)\nonumber \\
&+& \delta^{i}_{j}\tau^{-1}\sum_{l\neq j}\Pi_l(t)\left(pR^{i}_{l}(n^i_i-\rho_i^i(t))+(n^i_l-\rho^i_l(t))\right) \nonumber\\
&+& (1-\delta^{i}_{j})(1-\mu)(1-\tau^{-1})(pR^{i}_{j}\rho^i_i(t)+\rho^i_j(t))\nonumber \\
&+&  (1-\delta^{i}_{j})(1-\tau^{-1})\Pi_j(t)(pR^{i}_{j}(n^i_i-\rho^i_i(t))+(n^i_j-\rho^i_j(t)))\ .
\label{Eq.rho}
\end{eqnarray}
The first three lines resume the change in the infected agents located at their residence. Namely, first term gathers all the infected agents with residence in $i$ not recovering from the disease and staying at their destination whereas the rest involve contagions affecting susceptible individuals. This way, agents from $i$ can be infected inside their residence with probability $\Pi_i (t)$ or inside any other neighboring area $j$ with probability $\Pi_j$ and decide afterwards to come back home. Finally, fourth and fifth lines encode the evolution of infected agents located outside their residence. Analogously to the former case, they include those agents not overcoming the disease and not coming back home (fourth line), and those susceptible getting the disease at their destination with probability $\Pi_j (t)$ (fifth line). Following the well-mixing assumption for the interaction inside each patch, the probability of becoming infected inside a node $i$ at time $t$, $\Pi_i(t)$, reads:
\begin{equation}
\Pi_i(t) = 1-\prod\limits_{j=1}^{N}\left(1-\lambda\frac{(\rho^{int})^{j}_i(t)}{(n^{int})^{j}_i}\right)^{(n^{int})^{j}_i}\ ,
\end{equation}
where $(\rho^{int})^{j}_i(t)$ denotes the number of infected individuals with residence in $j$ and interacting at $i$ at time $t$. Similarly, $(n^{int})^{j}_i$ represents the total population from $j$ located at $i$ at the interaction stage. According to the mobility flows dictated by tensor ${\bf M}$ defined in Eq.(\ref{Eq.tensor}), both quantities are given by:
\begin{eqnarray}
(\rho^{int})^j_i (t)&=&\delta^{i}_{j}\left(1-p(1-R^{j}_{j})\right)\rho^j_j(t)+(1-\delta^{i}_{j})\left(\rho^j_i + pR^{j}_{i}\rho^j_j(t)\right) \\
(n^{int})^j_i &=&\delta^{i}_{j}\left(1-p(1-R^{j}_{j})\right)n^j_j+(1-\delta^{i}_{j})\left(n^j_i + pR^{j}_{i}n^j_j\right)\ .
\label{Eq.Intpop}
\end{eqnarray}

\subsection{Limiting cases of the Markovian equations}
Let us analyze the two limiting cases of the former equations given by $\tau^{-1}=1$ and $\tau^{-1}=0$. On the one hand, 
when $\tau^{-1}=0$ the information of residence is no longer relevant, since all the individuals are steadily redistributed across the metapopulation. In particular, the metapopulation turns into an effectively disconnected set of patches whose steady population is given by:
\begin{equation}
N^{st}_i=\sum\limits_{j=1}^{P} N^j R^{j}_{i}\, .
\end{equation}
On the other hand, the original MIR model can be recovered by setting the return rate to $\tau^{-1}=1$. In this scenario, Eq. (\ref{Eq.population}) yields:
\begin{equation}
n^i_j = \delta^{i}_{j}N^{i},
\end{equation}
which indicates that all the agents are at their residence when each time step starts. Thus, the Markovian model is reduced to $P$ equations, each one tracking the evolution of the fraction of infected agents associated to each patch, denoted as $\rho^i$. In particular, after normalizing by the population of each node, Eq. (\ref{Eq.rho}) turns into:
\begin{equation}
\rho^i(t+1)=(1-\mu)\rho^i(t) + (1- \rho^i(t))\left((1-p)\Pi_i(t)+p\sum\limits_{j=1}^{N} R^{i}_{j}\Pi_j\right),
\end{equation}
thus reproducing the MIR model introduced in \cite{natphys} to show that by increasing human mobility $p$ from $p=0$ the epidemic threshold $\lambda_c$ always increases, pointing out a non-trivial detrimental effect of mobility over the spread of diseases. In what follows, we validate the generalized Markovian equations and unveil the role that the new time scale $\tau$ plays on the aforementioned epidemic detriment driven by mobility.

\section{Validation}

To validate Eqs.(\ref{Eq.rho}-\ref{Eq.Intpop}), we generate a Barab\'asi-Albert synthetic metapopulation with $P=50$ patches and $\langle k \rangle=4$. Regarding agents distribution, we assume that the population of each patch, say $i$, is assigned according to a uniform distribution spanning the range $N_i\in[300,1700]$. We first study the evolution of the epidemic size as a function of both mobility and epidemiological parameters. Here we define epidemic size as the fraction of the whole population who remains infected once the epidemic has reached the stationary state. Theoretically, we compute this indicator by introducing an infectious seed, encoded in the formalism by a non-zero initial configuration $\left\lbrace \rho^i_j (0)\right\rbrace$, and iterating Eqs. (\ref{Eq.rho}-\ref{Eq.Intpop}) until no fluctuations in the epidemic size are observed. On the other hand, to carry out the mechanistic simulations, we introduce a small number of infected individuals and let the population evolve according to the microscopic rules defined in Sec. II. Due to the stochastic nature of mechanistic simulations, all the results are averaged over 100 realizations.  

\begin{figure}[t!]
\centering
\includegraphics[width=0.70\columnwidth,angle=-90]{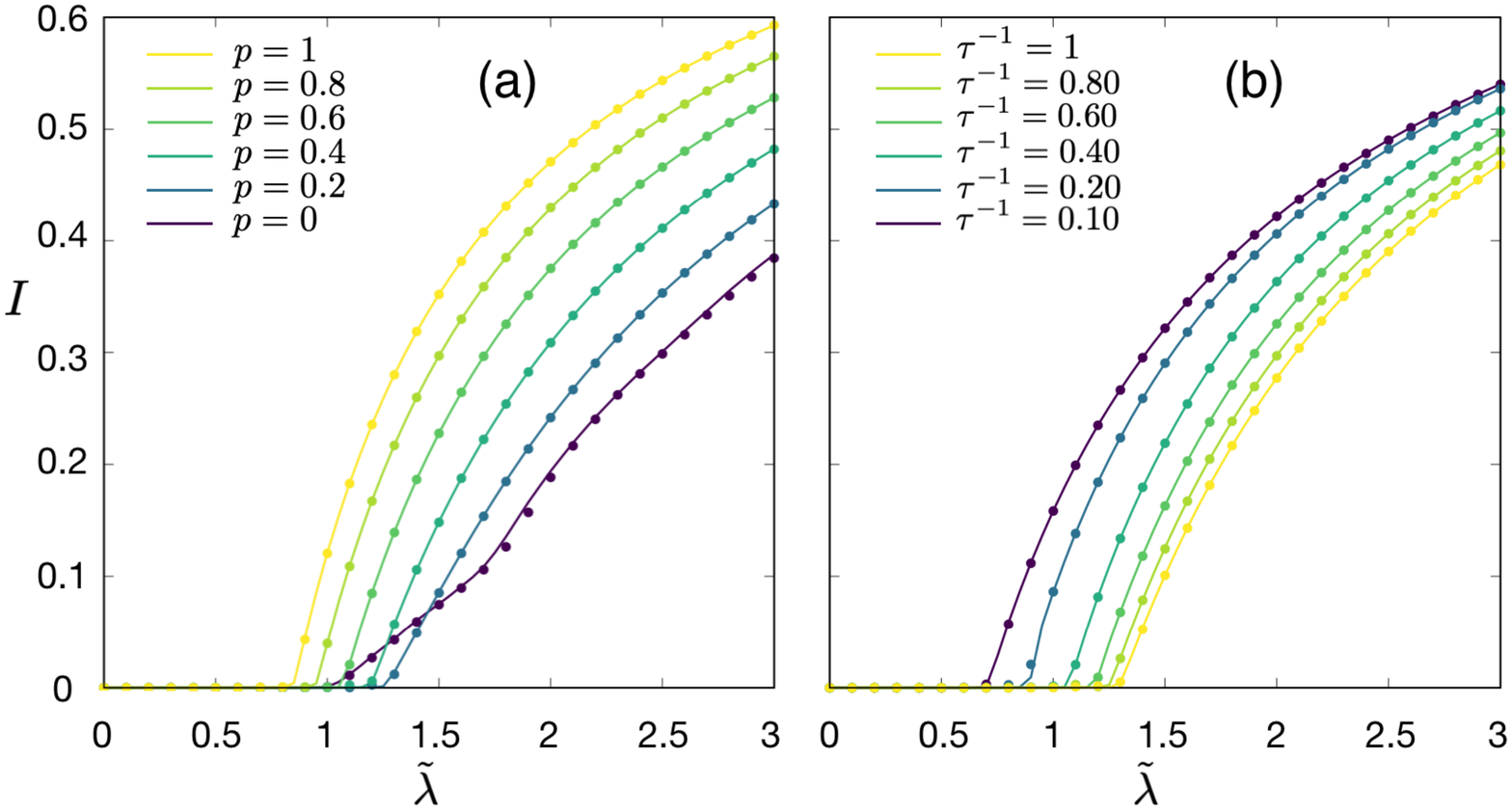}
\caption{Panel a): Epidemic size (see text for details) $I$ as a function of the infectivity $\tilde{\lambda}$ and the mobility rate $p$ (color code). The return rate has been set to $\tau^{-1}=0.5$. Panel b) Epidemic size (see text for details) $I$ as a function of the infectivity $\tilde{\lambda}$ and the return rate $\tau^{-1}$ (color code). The mobility rate has been set to $p=0.5$. In both panels, solid lines represent theoretical predictions obtained by iterating the Eqs. (\ref{Eq.rho}-\ref{Eq.Intpop}) whereas dots show results from Monte Carlo simulations averaged over 100 realizations. The recovery rate in both panels has been set to $\mu=0.2$. Note that the infectivity $\tilde{\lambda}$ has been re-scaled by the epidemic threshold for the static case, i.e, $\tilde{\lambda}=\lambda/\lambda_c(p=0)=\lambda N_{\textrm{max}}/\mu$.}
\end{figure}

Fig. 2 shows the great agreement between both theoretical (curves) and numerical (dots) results. In addition, we observe that the epidemic threshold exhibits a non-trivial dependence on both parameters $p$ and $\tau^{1}$. In particular, Fig. 2.a reveals that, keeping fixed the permanence time $\tau$ ($\tau=2$ in the figure),  the variation of the mobility rate $p$ leads to the emergence of a non-monotonous behavior of the epidemic threshold, as illustrated  in \cite{natphys} for the original MIR model ($\tau=1$). Namely, for low $p$ values, the epidemic threshold increases with the mobility, indicating that  mobility is detrimental to the spread of diseases. However, when $p$ is high, the dependence is inverted and the increase of human mobility leads to a decrease of the epidemic threshold, thus boosting disease spreading.  On the other hand, to analyze the influence of the temporal scale associated to movements, we set $p=0.5$ and study the effect of varying the return rate $\tau^{-1}$ in Fig. 2b. There, it becomes clear that by increasing the time associated to displacements fosters epidemic spreading since the epidemic threshold always decreases as $\tau$ increases.

\section{Analytical estimation of the epidemic threshold}
Here we aim at deriving an analytical expression of the epidemic threshold to capture the relevance of the different time scales for the onset of diseases. The epidemic threshold is defined as the minimum infectivity triggering epidemic outbreaks. To estimate its value, we make use of the next generation matrix method (NGM) \cite{NGM}, which enables to get a compact expression of the basic reproduction number $R_0$.

First, to apply the NGM technique we turn the discrete equations into continuous ones. Since we are interested in the epidemic threshold, we assume that $(\rho^{int})^{j}_i/(n^{int})^{j}_i=\epsilon^j_i \ll 1 \, \forall (i,j)$. This allows us to linearize Eqs. (\ref{Eq.rho}-\ref{Eq.Intpop}) by neglecting terms $\mathcal{O}(\epsilon^2)$. For the sake of readability, let us now explicitly express the evolution of infected agents inside and outside their residence. Once linearized the equations, the temporal dependence of the number of infected agents inside their residence $\dot{\rho}^i_i$ reads:
\begin{eqnarray}
\dot{\rho}^i_i &=&\underbrace{-\mu\rho^i_i-p(1-R^{i}_{i})(1-\mu)(1-\tau^{-1})\rho^i_i+\sum\limits_{l\neq i}(1-\mu)\tau^{-1}\rho^i_l}_V \nonumber \\
&+& \lambda\underbrace{\sum\limits_l(1-p(1-R^{i}_{i}))n^i_i\left[\delta^{l}_{i}(1-p(1-R^{i}_{i})\rho^i_i + (1-\delta^{l}_{i})(pR^{l}_{i}\rho_l^l+\rho^l_i)\right]}_F\nonumber \\
&+& \lambda\underbrace{\tau^{-1}\sum\limits_m\sum\limits_{l\neq i} \left(pR^{i}_{l}n^i_i + n^i_l\right)\left[\delta^{l}_{m}\left[(1-p(1-R^{l}_{l}))\rho^l_l\right]+(1-\delta^{l}_{m})\left[pR^{m}_{l}\rho^m_m+\rho^m_l\right]\right]}_F\ .\nonumber\\ 
\label{Eq.ii}
\end{eqnarray}
In case that $i\neq j$, the former temporal evolution is given by:
\begin{eqnarray}
\dot{\rho}^i_j&=&\underbrace{-\mu\rho^i_j-\tau^{-1}(1-\mu)\rho^i_j + (1-\mu)(1-\tau^{-1})pR^{i}_{j}\rho^i_i} _V\nonumber \\
&+& \lambda\underbrace{(1-\tau^{-1})(pR^{i}_{j}n^i_i+n^i_j)\sum\limits_l\left(\delta^{j}_{l}\left[1-p(1-R^{l}_{l})\right]\rho^j_j + (1-\delta^{j}_{l})\left[pR^{l}_{j}\rho^l_l + \rho^l_j\right]\right)}_F \nonumber\\ 
\label{Eq.ij} 
\end{eqnarray}
Expressions (\ref{Eq.ii}-\ref{Eq.ij}) can be rewritten in a more compact way by defining two 4-dimensional tensors ${\bf F}$ and ${\bf V}$. This way, the evolution of the number of infected individuals with residence in $i$ located at $j$ can be generally expressed as:
\begin{equation}
\dot{\rho}^i_j= \left(\lambda F^{il}_{jm}-V^{il}_{jm}\right)\rho^m_l\ .
\end{equation}
Note that tensor ${\bf F}$ contains all the processes giving rise to an increase of the epidemic size whereas tensor ${\bf V}$ encodes the rest of transitions affecting infected agents. From the latter,  the basic reproductive number $R_0$ can be obtained as:

\begin{equation}
R_0 = \frac{1}{\lambda\Lambda_{\textrm{max}}({\bf FV^{-1}})}\ ,
\end{equation}
being $\Lambda_{\textrm{max}}({\bf FV^{-1}})$ the spectral radius of the matrix ${\bf FV^{-1}}$.  Recalling that the epidemic threshold corresponds to the value of $\lambda$ leading to $R_0=1$, $\lambda_c$ reads as:
\begin{equation}
\lambda_c =  \frac{1}{\Lambda_{\textrm{max}}({\bf FV^{-1}})}\ .
\label{Eq.lambdac}
\end{equation}

\begin{figure}[t!]
\centering
\includegraphics[width=0.90\columnwidth]{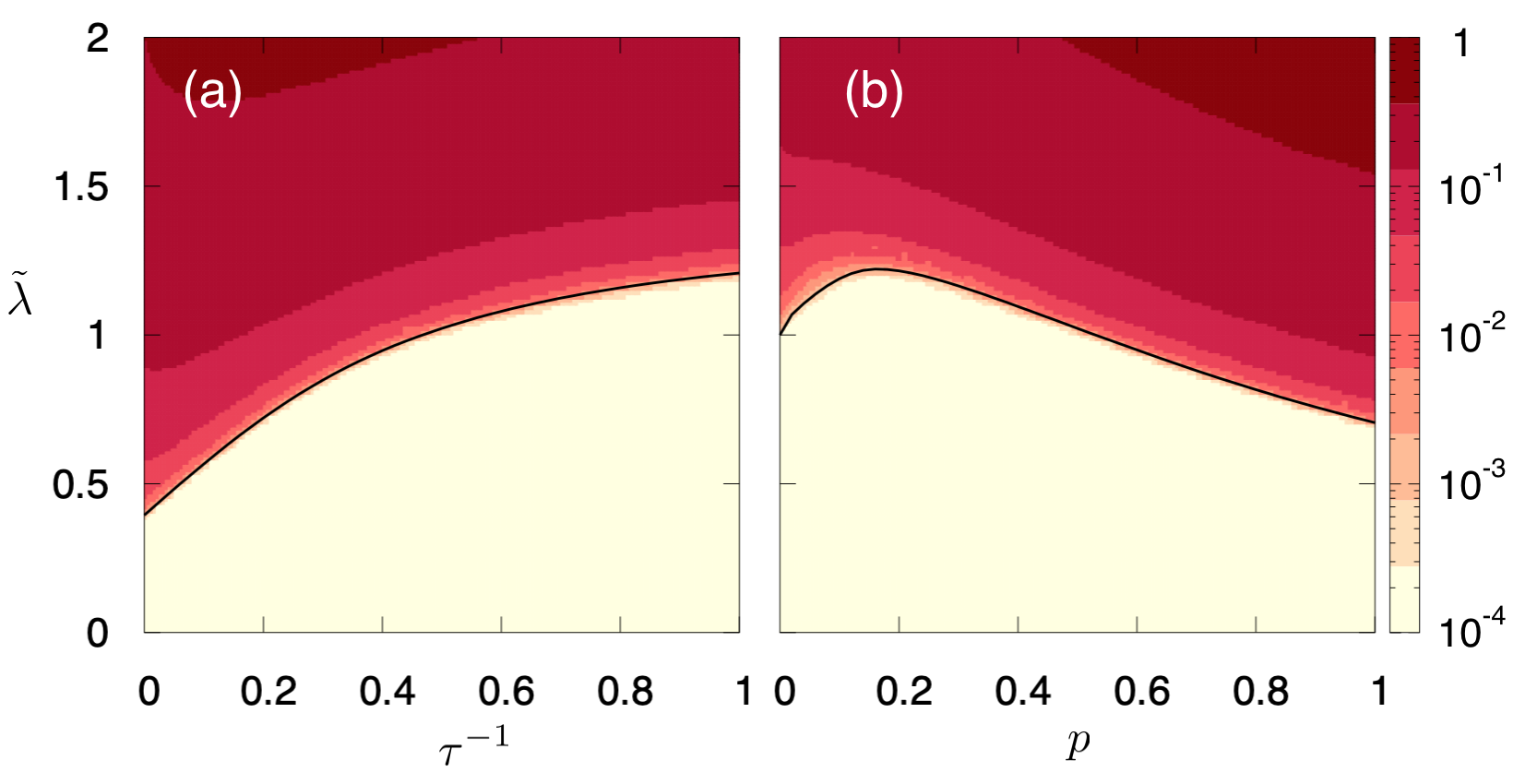}
\caption{Left panel: Epidemic size (color code) as a function of the infectivity $\tilde{\lambda}$ and the probability of returning to home $\tau^{-1}$. The mobility rate $p$ has been set to $p=0.5$. Right panel:  Epidemic size (color code) as a function of the infectivity $\tilde{\lambda}$ and the mobility rate $p$. The return rate has been set to $\tau^{-1}=0.50$. In both panels, the solid black line shows the estimation for the epidemic threshold $\lambda_c(p,\tau^{-1})$ computed from Eq. (\ref{Eq.lambdac}). The recovery rate has been fixed to $\mu=0.2$.}
\label{Fig.ValThreshold}
\end{figure}

To study the effects of tuning human mobility on the epidemic threshold, we consider the dependence of the normalized epidemic threshold $\tilde{\lambda}$ on both the degree of mobility $p$ and the probability of returning home $\tau^{-1}$. We define $\tilde{\lambda}$ as the epidemic threshold re-scaled by the epidemic threshold corresponding to the static case, {\em i.e.}, $\tilde{\lambda}(p,\tau^{-1})=\lambda_c(p,\tau^{-1})/\lambda_c(0,\tau^{-1})$. Fig.~\ref{Fig.ValThreshold} reveals the great accuracy of Eq.~(\ref{Eq.lambdac}) for capturing the boundary between the disease-free region and the epidemic state. In addition, our theoretical estimation is also able to reproduce the non-monotonous impact of human mobility on the epidemic threshold and the promotion of epidemic spreading while increasing the permanence times at the destination.

\section{Interplay between mobility and epidemiological time scales}

In the previous section, we have obtained a theoretical expression of the epidemic threshold which enables to characterize the influence of human mobility on the onset of diseases. Nonetheless, the computation of this value involves solving an eigenvalue problem of a $P^2\times P^2$ matrix, which clearly limits getting any insights on the role that each parameter plays. We are now interested in finding simple heuristic arguments which allows us to better understand the physical roots of the observed behavior of the epidemic threshold. 

Let us start by analyzing the static scenario in which nobody moves ($p=0$) so all the agents only interact with the other residents in their associated patch. In this particular case, each patch displays a different epidemic threshold $\lambda^i_c$ given by\cite{natphys}:
\begin{equation}
\lambda^i_c=\frac{\mu}{N_i}\ .
\end{equation}
Therefore, the global epidemic threshold is readily obtained for the static case as:
\begin{equation}
\lambda_c(p=0,\tau^{-1})=\frac{\mu}{N_{\textrm{max}}}\ ,
\end{equation}
being $N_{\textrm{max}}$ the number of residents in the most populated patch. 

\begin{figure}[t!]
\centering
\includegraphics[width=1\columnwidth]{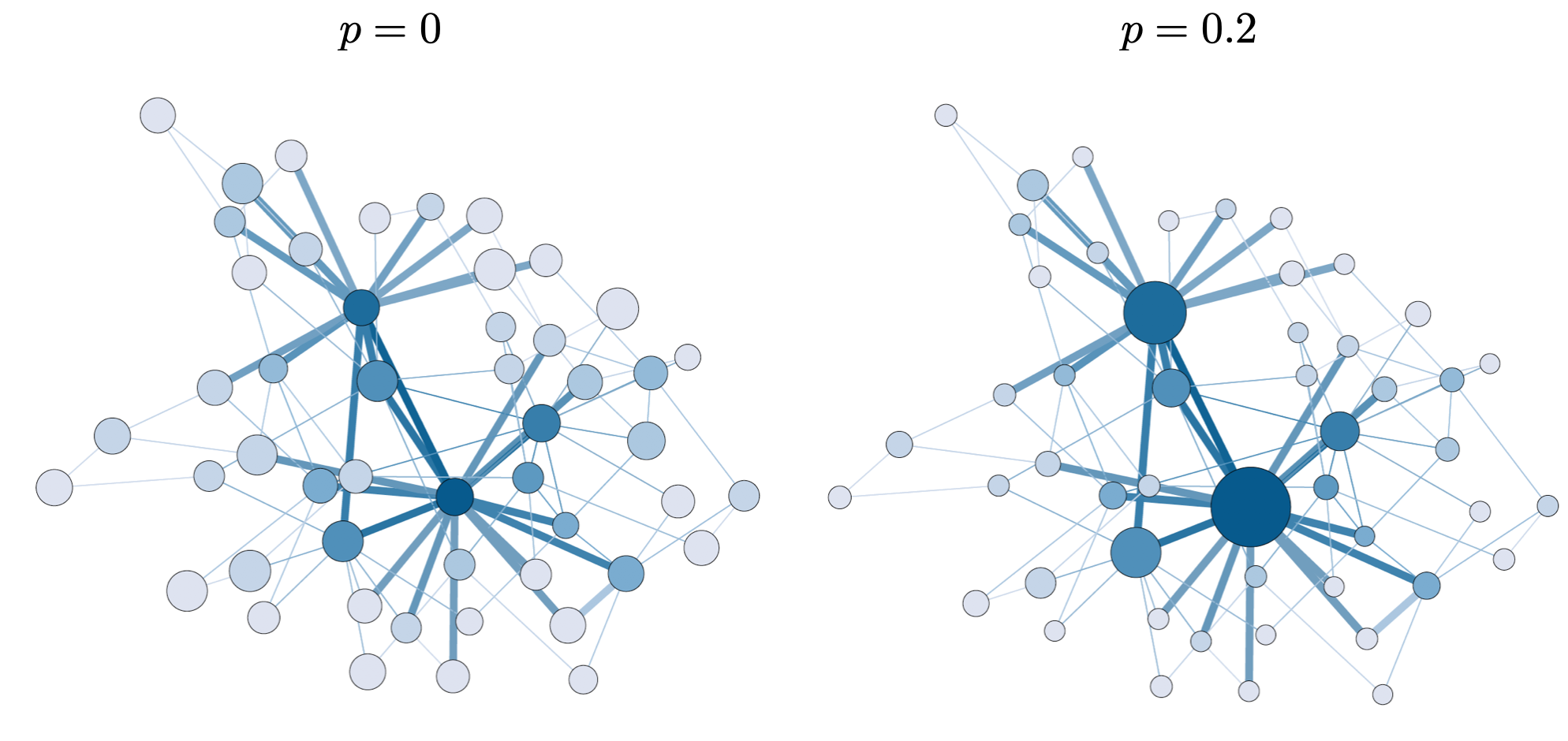}
\caption{Barab\'asi-Albert metapopulation governing human flows. The number of incoming connections of each patch is encoded in the color code ranging from the darker tones associated to highly connected patches to the brighter ones corresponding to barely connected areas. The size of the patches is proportional to the effective population interacting inside them. The mobility rate has been set to $p=0$ (Left) and $p=0.2$ (Right). The return rate is fixed to $\tau^{-1}=0.5$ in right panel. The number of residents in each area, say $i$, is assigned according to an uniform distribution within the range $N^{i}\in [300,1700]$.}
\label{Fig.SFrep}
\end{figure}

To unveil the role of the degree of mobility $p$, we fix the return rate to $\tau^{-1}=0.5$ and represent in Fig. 4 the mobility network along with the population distribution for the static case and for a non-zero value of the mobility. There we observe how human mobility clearly promotes the accumulation of agents inside the most connected nodes.  Following our previous arguments, one could intuitively argue that the increase of mobility decreases the epidemic threshold due to the promotion of crowded areas that foster the spread of diseases. However, to understand the epidemic detriment observed in Fig.~{\ref{Fig.ValThreshold}.a, we should also consider that, at the individual level, increasing human mobility also facilitates the spatial diffusion of infected agents, thus preventing their fixation inside those crowded subpopulations during their whole infectious period (of typical duration $\mu^{-1}$ time steps). Therefore, mobility promotes epidemic spreading because of the accumulation of agents in some areas but also hinders it by keeping infected agents away from vulnerable areas. As a consequence, the observed value of the epidemic threshold for each value of the mobility $p$ is the outcome of the competition between such opposite effects. 

Interestingly, this trade-off is highly influenced by the return rate $\tau^{-1}$. In particular, the increase of the permanence time that visitors spend at their destination, {\em i.e.}, decreasing $\tau^{-1}$, reduces the beneficial effect of the mobility, for it implies that infected agents have a higher probability of being in the most affected areas during their typical infectious time, $\mu^{-1}$, thus promoting a larger number of contagions. 
\begin{figure}[t!]
\centering
\includegraphics[width=1\columnwidth]{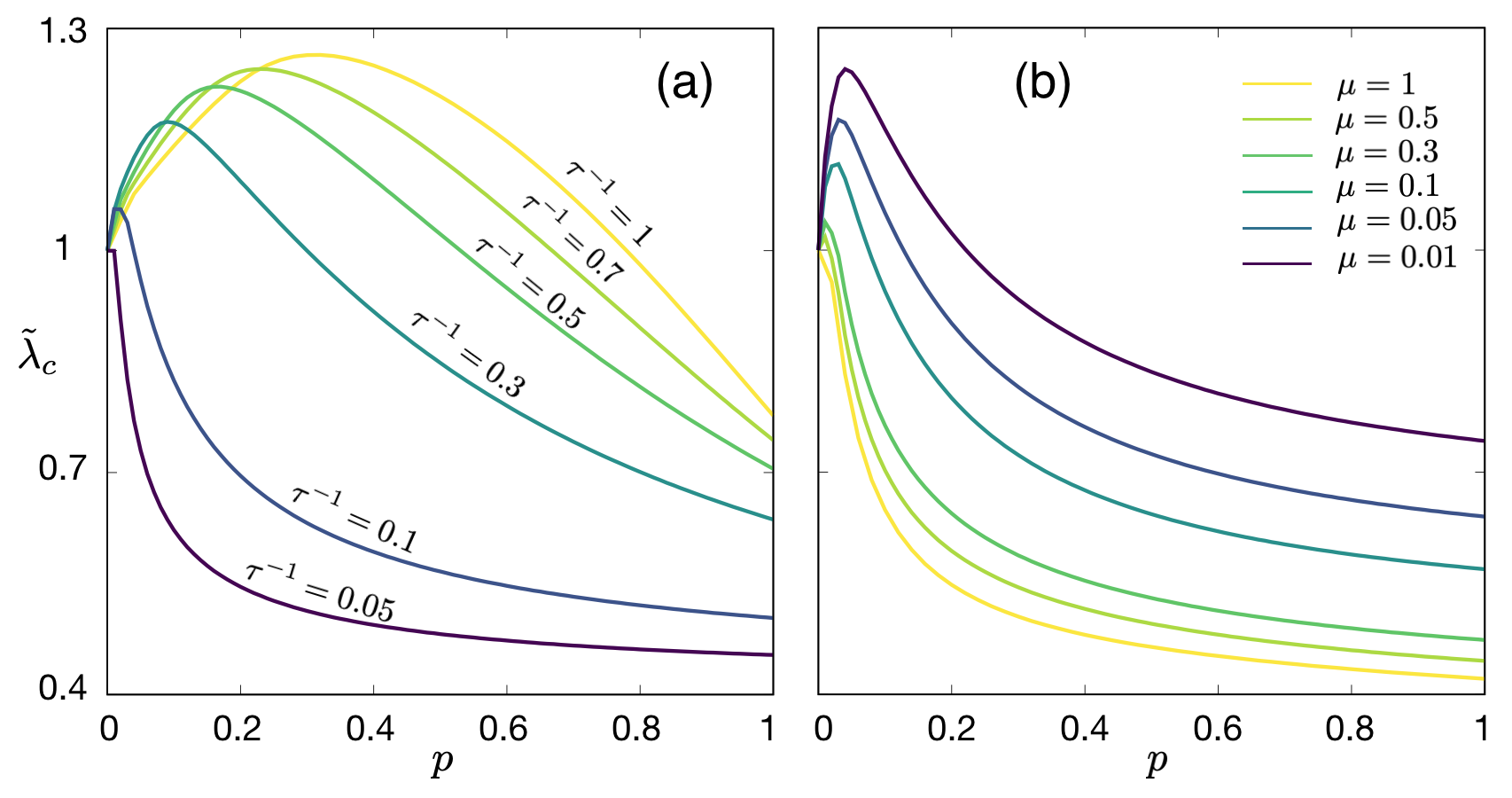}
\caption{Left panel: Normalized epidemic threshold $\tilde{\lambda}_c$ as a function of the mobility for several values of the return rate $\tau^{-1}$ (color code). The recovery rate has been set to $\mu=0.2$. Right panel: $\tilde{\lambda}_c$ as a function of the mobility for several values of the recovery rate $\mu$. The return rate is fixed to $\tau^{-1}=0.10$.}
\label{Fig.lambdataumu}
\end{figure}
To confirm this phenomenon, we represent in Fig.~\ref{Fig.lambdataumu}.a the dependence of the normalized epidemic threshold $\tilde{\lambda}_c$ on the mobility $p$ when tuning the permanence time $\tau^{-1}$. From this plot it becomes clear that, for small time scales associated to human movements (high values of $\tau^{-1}$), the spatial diffusion of infected agents can compensate the accumulation of agents due to the mobility. Therefore, the overall result is a detriment of the epidemic spreading when increasing $p$. In contrast, for long permanence times, the dissemination of infected agents is clearly reduced and the creation of crowded sources of infection gains relevance, thus leading to the enhancement of epidemic spreading with mobility.

\begin{figure}[t!]
\centering
\includegraphics[width=1\columnwidth]{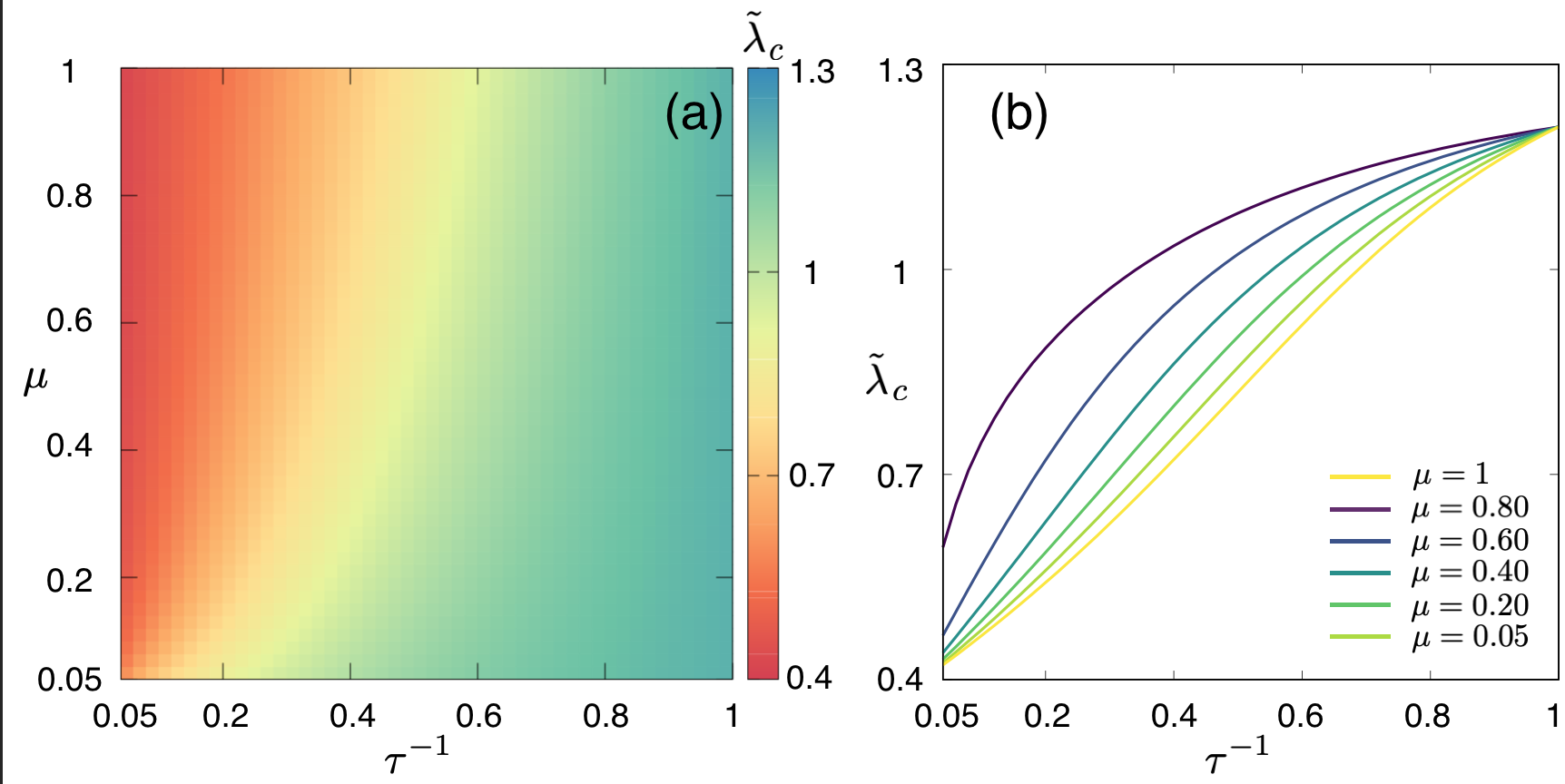}
\caption{Left panel: Normalized epidemic threshold $\tilde{\lambda}_c$ (color code) as a function of the recovery rate $\mu$ and the return rate $\tau^{-1}$. Right panel: Dependence of the normalized epidemic threshold on the return rate $\tau^{-1}$ for several values of the recovery rate $\mu$. The mobility rate has been fixed in all panels to $p=0.5$.}
\end{figure}

It is is also worth studying the effects of the interplay between both epidemiological and mobility time scales in shaping epidemic outbreaks. To do so, we fix the return rate to $\tau^{-1}=0.1$ and represent (see in Fig.~\ref{Fig.lambdataumu}.b) the normalized epidemic threshold $\tilde{\lambda}_c$ as a function of the mobility $p$ for several values of the recovery rate $\mu$. From this plot it becomes clear that increasing the typical recovery time ($\mu^{-1}$) suppresses the effect of introducing larger time scales associated to human movements since epidemic detriment is observed only when $\mu<\tau^{-1}$. This is an expected result since, following our previous arguments, as long as the typical epidemiological time scale $\mu^{-1}$ is much longer than the one associated to movements, $\tau$, mobility allows an infected individual to visit different areas far from crowded subpopulations before overcoming the disease. In contrast, when $\mu^{1}<\tau$, infected agents spend most of their epidemiological cycle inside one specific area. In this sense, the creation of contagions sources due to human mobility along with the continuous presence of infected agents there lead to the decrease of the epidemic threshold when promoting human movements.

Finally, we shed more light into the antagonistic character of the epidemic time scale $\mu^{-1}$ and the permanence time of human movements $\tau$. For this purpose, we study the surface $\tilde{\lambda} (\mu,\tau^{-1})$ while keeping fixed the degree of mobility to $p=0.5$ in Fig.~6.a. There we observe that, as stated above, the increase of the time scale governing recovery processes (decrease of $\mu$) gives rise to an increase of the normalized epidemic threshold $\tilde{\lambda}_c$, thus diminishing the effect of prolonging the permanence times at the contagion focus. This effect  becomes more evident in Fig.~6.b where we explicitly show the dependence of the normalized epidemic threshold on the inverse of the typical permanence time, $\tau^{-1}$. It can be noticed how the increase of the permanence time ({\em i.e.} reducing $\tau^{-1}$) always boosts the spread of the disease by reducing the epidemic threshold. This effect is much more accused when the epidemic time scale is smaller than the typical trip duration.

\section{Conclusions}
In this manuscript we have studied the influence of recurrent human mobility patterns on the onset of epidemics at different temporal scales. The main contribution of this work is to provide a metapopulation framework to obtain theoretical insights about the role of different of some parameters related to human mobility, extending and complementing previous results \cite{natphys, multmetapop1}. Importantly, this framework allows to incorporate data about the population distribution (demography) and the back-and-forth mobility patterns from Origin-Destination matrices of real populations such as cities, regions, countries, etc. 
The presented framework is formulated through a set of Markovian evolution equations that generalize those introduced in \cite{natphys} by incorporating the time scale associated to the duration of trips, here represented by the permanence time in the destination nodes. From a computational point of view, the Markovian framework introduced in this work constitutes a computational time-saving approach since the accuracy of the theoretical predictions provided by their equations allows to monitor the spatio-temporal evolution of SIS diseases without the necessity of carrying out lengthy individual-based mechanistic simulations. 

Moreover, we have obtained a theoretical expression of the epidemic threshold, which allows us to determine the critical conditions triggering the onset of epidemics. Interestingly, we have revealed that the design of policies related to human mobility  to tackle epidemic outbreaks is closely related to the interplay between the different temporal scales associated to human movements and epidemic processes. In this sense, we have shown that the dependence of the epidemic threshold on some mobility related parameters such as the frequency of trips or their duration is strongly shaped by the typical recovery rate of the disease. This way, the promotion of containment measures regarding mobility should be focused on each particular disease to efficiently reduce their impact on society. 

This work helps to bridge the gap between metapopulation models and real epidemic scenarios for which information about the demographical distribution of individuals and their mobility patterns is available. Notwithstanding, some of the assumptions  are still far from being realistic. In this sense, our formalism paves the way to the elaboration of more realistic frameworks by considering other contagion mechanisms like those associated to vector-borne diseases \cite{vbd1,vbd2}, or by including more realistic scenarios regarding mobility such as the multiplex nature of human flows \cite{multiplex1,multiplex2,multiplex3}, the existence of mobility patterns beyond recurrent movements \cite{traject1,traject2,traject3} or the modification of human mobility patterns as a response to epidemic outbreaks \cite{epi3,dynamicmobility}.

\section*{Acknowledgments}
D.S.-P. acknowledges financial support from Gobierno de Aragon through a doctoral fellowship. D.S.-P and J.G.G. acknowledge financial support from Ministerio de Econom\'{\i}a y Competitividad (grants FIS2015-71582-C2 and FIS2017-87519-P) and Gobierno de Arag\'on with Fondo Social Europeo (grant E36-17R). AA acknowledges support by Ministerio de Econom\'{\i}a y Competitividad (grant FIS2015-71582-C2-1), Generalitat de Catalunya (grant 2017SGR896), Universitat Rovira i Virgili (grant 2017PFRURV-B2-41), ICREA Academia and the James S. McDonnell Foundation (grant 220020325).


\begin{thebibliography}{99}
\bibitem{nota} Note that, according to the our notation, superindexes denote the residence patch whereas subindexes account for the nodes where agents are currently placed and/or move to.
\end{thebibliography}
\end{document}